\documentclass[10pt,aps,twocolumn,nofootinbib,superscriptaddress,eqsecnum]{revtex4-2}
\usepackage[T1]{fontenc}
\usepackage[utf8]{inputenc}
\usepackage{orcidlink}
\usepackage{nicematrix}
\usepackage{amsmath,amssymb,amsthm,bm,mathrsfs}
\usepackage[normalem]{ulem}
\usepackage{hyperref}
\usepackage{tikz}
\usetikzlibrary{arrows.meta,positioning}
\newtheorem{theorem}{Theorem}
\newcommand{\Tr}{\mathrm{Tr}}
\allowdisplaybreaks
\begin{document}
\newcommand{\Corrado}[1]{\textcolor{red}{#1}}
\newcommand{\resizeeq}[1]{%
  $\vcenter{\hbox{\rule{0pt}{0pt}}\vskip 1pt
            \hbox{\resizebox{0.15\columnwidth}{!}{#1}}%
            \vskip 1pt}$%
}
\title{Third-Order Quantum Information from Distributed Randomized Measurements}
\author{Giovanni Scala~\orcidlink{0000-0003-2685-0946}}
	\affiliation{Institute of Physics, Faculty of Physics, Astronomy and Informatics, Nicolaus Copernicus University, Grudziądzka 5/7, 87-100 Toruń, Poland}
    
\author{Corrado C. Mattiacci~\orcidlink{0009-0005-4554-5260}}
    \affiliation{Institute of Physics, Faculty of Physics, Astronomy and Informatics, Nicolaus Copernicus University, Grudziądzka 5/7, 87-100 Toruń, Poland}

\author{Gniewomir Sarbicki~\orcidlink{0000-0002-6403-8839}}
    \affiliation{Institute of Physics, Faculty of Physics, Astronomy and Informatics, Nicolaus Copernicus University, Grudziądzka 5/7, 87-100 Toruń, Poland}
    \affiliation{Institute of Advanced Studies, Nicolaus Copernicus University in Toruń, Wileńska 4, 87-100 Toruń, Poland}

\begin{abstract}
Randomized measurements offer a way to probe nonlinear properties of quantum
states without reconstructing their full density matrices.
For states shared between distant laboratories, we study products of three
outcome probabilities measured at the same local settings and then averaged
over independent local random unitaries. We determine, for an arbitrary
number of parties, which third-order local-unitary invariant combinations
these averages retain and which they cannot separate.
The accessible information is encoded in algebraic invariants of the quantum state and is reconstructed by analysing the permutation symmetries induced by the distributed measurement protocol. Exploiting these symmetries, the reconstruction problem collapses, regardless of the number of measurement outcomes, to the symbolic inversion of a $3\times3$ matrix. As a result, nonlinear information can be extracted without inverting the exponentially large matrices that typically arise in distributed multipartite quantum systems.
\end{abstract}
\maketitle

\section{Introduction}\label{sec:introduction}

Many properties relevant to quantum-state characterization are nonlinear
functions of the density matrix, yet can be determined without
full-state reconstruction~\cite{vanEnk2012,Ekert2002,Brun2004,Leifer2004,Elben2023toolbox,cieslinski2024analysing}.
Randomized measurements provide access to such information through
products of outcome probabilities averaged over random measurement
bases~\cite{elben2019statistical,Elben2018renyi,Brydges2019}.
For spatially separated subsystems, the central question is which
nonlinear properties can be recovered from local measurements and the
classical records shared between laboratories.

We study third-order correlators obtained by measuring three independent
preparations at the \emph{same} local unitary setting and averaging over Haar distribution. The protocol requires only local
rotations, projective local measurements, and classical post-processing,
with no global measurement or coherent coupling between copies.
After averaging, each local outcome triple is specified by its equality
pattern: all distinct, exactly one equal pair, or all equal.
Replacing repeated averaged entries by these patterns loses no further
information. The local Haar averages assign identical coefficients
to some distinct invariants, so they appear in the measurement equations
only through sums. We ask which invariant combinations these equations
determine and which they cannot separate.

Classical-shadow protocols use recorded settings and outcomes to
estimate selected observables~\cite{Huang2020}.
Elben et al.~\cite{Elben2020} combine records from independently
sampled settings to estimate partial-transpose moments individually.
Other randomized-measurement schemes access additional invariant
polynomials or optimize estimation costs~\cite{Liu2022permutation,
WyderkaKetterer2023,Rath2021importance,Vermersch2024}.

Our reconstruction takes only the averaged outcome
correlators as input. Unlike estimators that use recorded settings,
it requires neither calibration nor storage of each individual unitary
matrix: one only needs to identify shots taken at the same setting.
This removes the need to characterize every random rotation for
post-processing, while retaining the assumptions of Haar sampling
(or an exact \(3\)-design) and the specified measurement model.
For resource tests built from the retained invariant combinations,
only the combinations required by the test need to be reconstructed.

We solve this identifiability and reconstruction problem for an arbitrary
number of parties and local dimensions. Simultaneous relabelling of the three copies organises
the retained coordinates into permutation orbits. A change of basis,
independent of the local Hilbert-space dimensions, then separates the
reconstruction map into independent sectors built from $3\times 3$
matrices and scalars.

The construction recovers the known bipartite reconstruction
structure~\cite{QST2025,scala2026finite}, including the fact that the
global and partial-transpose third moments enter only through their sum,
and extends that structure to arbitrary party number.
Its contribution is an exact characterization of the nonlinear
information available from distributed measurements.

Section~\ref{sec:experiment} defines the protocol.
Sections~\ref{sec:one_party} and \ref{sec:two_parties} develop the
one- and two-party constructions, and Sec.~\ref{sec:general} establishes
their multipartite extension.
Section~\ref{sec:sectors} constructs the reconstruction sectors, followed
by the main theorem in Sec.~\ref{sec:theorem}.
Section~\ref{sec:blindness} then examines partial-transpose invariance
and the information limits of the averaged correlators.
\section{Randomized measurements}\label{sec:experiment}
Let \(\rho\) be a density matrix on
\(\bigotimes_{k=1}^n\mathbb C^{d_k}\).
Party $k$ applies a unitary $U_k$ and measures in the computational basis $\{|i_k\rangle\}_{i_k=1}^{d_k}$. Write $\boldsymbol U=(U_1,\ldots,U_n)$ for the setting and $\boldsymbol i=(i_1,\ldots,i_n)$ for a joint outcome tuple. The conditional probability is
\begin{equation}
 p_\rho(\boldsymbol i|\boldsymbol U)
 =\operatorname{Tr}\!\left[\rho\bigotimes_{k=1}^n
 U_k^\dagger|i_k\rangle\langle i_k|U_k\right].
 \label{eq:born}
\end{equation}
For three independently prepared copies measured at the same product setting
\(\boldsymbol U\), the joint conditional probability of the outcome tuples
\(\boldsymbol i_1,\boldsymbol i_2,\boldsymbol i_3\) factorizes as
\begin{equation}
 p_\rho(\boldsymbol i_1,\boldsymbol i_2,\boldsymbol i_3|\boldsymbol U)
 =
 \prod_{h=1}^3 p_\rho(\boldsymbol i_h|\boldsymbol U).
 \label{eq:product}
\end{equation}
Each probability is linear in \(\rho\), so their product is
cubic. The three shots use the \emph{same setting}, and then their product is averaged
over independent local Haar distributions,
\begin{align}
 &y_{\rm out}(\boldsymbol i_1,\boldsymbol i_2,\boldsymbol i_3)
 =\int\prod_{k=1}^n\mathrm dU_k\,
 \prod_{h=1}^3p_\rho(\boldsymbol i_h|\boldsymbol U).\label{eq:haar_data}\\
 &=
\sum_{\boldsymbol q_h,\boldsymbol p_h}
\left(\prod_{h=1}^3
\rho_{\boldsymbol q_h,\boldsymbol p_h}\right)
\prod_{k=1}^n
\int dU_k
\prod_{h=1}^3
(U_k)_{i_h^kq_h^k}
\overline{(U_k)}_{i_h^kp_h^k}.
\nonumber
\end{align}
This probes third-order
information without a coherent measurement coupling the copies.
Collecting all outcome triples gives the data vector
\(\boldsymbol y_{\rm out}^{(n)}\in\mathbb R^{\prod_k d_k^3}\).
The integrand contains three matrix entries of each local unitary
and three conjugate entries. An exact unitary \(3\)-design, an
ensemble matching these Haar moments, therefore gives the same mean
correlators~\cite{GrossAudenaertEisert2007,Mele2024}.
For a qubit register at each party, the multiqubit Clifford group
provides such an ensemble~\cite{Zhu2017}.
We first develop the construction for $n=1$ to
introduce the building blocks for the main theorem. 
\section{One party}\label{sec:one_party}

\subsection{Equality patterns and local cycle merging}

For one party, the Haar average in Eq.~\eqref{eq:haar_data}
connects measured outcome triples to scalar invariants of \(\rho\).
Weingarten calculus evaluates the required integral as a sum over
permutations of the three copy positions~\cite{Collins2006,Collins2003,
CollinsMatsumotoNovak2022}:
\begin{align}
 &\int\mathrm dU\prod_{h=1}^3
 U_{i_hq_h}\overline{U}_{i_hp_h}
 \nonumber\\
 &\quad=
 \sum_{\sigma,\tau\in S_3}
 \mathrm{Wg}(\sigma\tau^{-1},d)
 \prod_{h=1}^3
 \delta_{i_h,i_{\sigma(h)}}\delta_{q_h,p_{\tau(h)}},
\end{align}
and the formula (\ref{eq:haar_data}) takes the form:
\begin{align}
 &y_{\rm out} (i_1,i_2,i_3) = 
 \sum_{\sigma,\tau\in S_3}
 \prod_{h=1}^3\delta_{i_h,i_{\sigma(h)}}
 \mathrm{Wg}(\sigma\tau^{-1},d)
 \nonumber \\
 &
 \sum_{p_1, p_2, p_3, q_1, q_2, q_3} \rho_{q_1 p_1} \rho_{q_2 p_2} \rho_{q_3 p_3} 
 \prod_{h=1}^3\delta_{q_h,p_{\tau(h)}}.
 \label{eq:weingarten}
\end{align}
Here \(\sigma\) tests equalities among the measured outcomes, whereas
\(\tau\) connects indices of the three density-matrix factors.
These two roles define the data and invariant coordinates below.

\paragraph{Equality patterns.}
The factor
\(\prod_{h=1}^3\delta_{i_h,i_{\sigma(h)}}=1\) iff all outcomes linked by \(\sigma\) are equal,
 e.g. \(\sigma=(12)\) requires\(i_1=i_2\), whereas \(\sigma=(123)\)
  requires \(i_1=i_2=i_3\).
Consequently, the averaged entry depends only on the equality
pattern of the outcome triple. We use the ordered set
\begin{equation}
 \Pi=(D,T_{12},T_{23},T_{13},E),
 \label{eq:patterns}
\end{equation}
where \(D\) denotes three distinct outcomes, \(T_{ab}\) denotes
equality only between positions \(a,b\), and \(E\) denotes three
equal outcomes.

For example, \((1,1,2)\) and \((2,2,3)\) both have pattern
\(T_{12}\) and therefore the same averaged entry \(y_{T_{12}}\).
The vector \(\boldsymbol y_{\rm patt}\in\mathbb R^5\) collects
one such value per pattern; its entries are representative
correlators, rather than total pattern-event probabilities.

Evaluating the outcome equality tests on each pattern defines
the matrix \(S_{\pi\sigma}\). With permutation columns ordered as
\((1,(12),(23),(13),(123),(132))\), it is
\begin{equation}
S :=
\begin{bNiceArray}[first-row]{c|ccc|cc}
1 & (12) & (23) & (13) & (123) & (132) \\
1 & 0 & 0 & 0 & 0 & 0 \\
\hline
1 & 1 & 0 & 0 & 0 & 0 \\
1 & 0 & 1 & 0 & 0 & 0 \\
1 & 0 & 0 & 1 & 0 & 0 \\
\hline
1 & 1 & 1 & 1 & 1 & 1 \\
\end{bNiceArray}
\begin{array}{c}
         D\\T_{12}\\T_{23}\\T_{13}\\E
    \end{array}
    \label{eq:S}
\end{equation}
Thus \(S\) encodes which permutation constraints each equality
pattern satisfies.

\paragraph{Permutation contractions.}
The index connections specified by \(\tau\) can be expressed
through the unitary permutation operator
\(P_\tau:(\mathbb C^d)^{\otimes3}\to(\mathbb C^d)^{\otimes3}\),
defined by
\begin{equation}
 P_\tau|i_1,i_2,i_3\rangle
 =
 |i_{\tau^{-1}(1)},i_{\tau^{-1}(2)},i_{\tau^{-1}(3)}\rangle.
 \label{eq:permutation}
\end{equation}
This convention gives \(P_\sigma P_\tau=P_{\sigma\tau}\).
The scalar obtained by closing the index connections is
\begin{equation}
 x(\tau)=\operatorname{Tr}[P_\tau\rho^{\otimes3}],
 \qquad
 \boldsymbol x_{\rm full}
 =(x(\tau))_{\tau\in S_3}\in\mathbb C^6.
 \label{eq:contractions_one}
\end{equation}
These contractions are invariant under unitary changes of basis
because \(P_\tau\) commutes with \(U^{\otimes3}\)
~\cite{Grassl1998,Rains2000,Makhlin2002,Kempe1999,
LindenPopescuSudbery1999,Vrana2011}.

For one party, the cycle structure of \(\tau\) determines which
state factors are joined. The identity, a transposition, and a
three-cycle give, respectively,
$q_0=(\operatorname{Tr}\rho)^3$,
 $q_1=(\operatorname{Tr}\rho)\operatorname{Tr}\rho^2$,
 $q_2=\operatorname{Tr}\rho^3$.
Accordingly,
\begin{equation}
 \boldsymbol x_{\rm full}
 =(q_0,q_1,q_1,q_1,q_2,q_2)^T.
 \label{eq:moments_full}
\end{equation}
The six labelled coordinates therefore contain only three
distinct moment values.

\paragraph{Local cycle merging.}
Let \(W_{\sigma\tau}(d)=\mathrm{Wg}(\sigma\tau^{-1},d)\).
It assigns to a $\sigma$, permitted by the equality pattern by $S$, a linear combination of unitary invariants encoded by $\tau$.
It depends only on whether its permutation argument is the identity $\textit{[1]}$, a transposition $\textit{[12]}$, or a three-cycle $\textit{[123]}$:
\begin{align}
\label{eq:Wg_values_e}
\mathrm{Wg}(\textit{[1]},d)&=\frac{d^2-2}{d(d^2-1)(d^2-4)} =: e,\\
\label{eq:Wg_values_t}
\mathrm{Wg}(\textit{[12]}, d)&=\frac{-d}{d(d^2-1)(d^2-4)} =: t,\\
\label{eq:Wg_values_c}
\mathrm{Wg}(\textit{[123]}, d)&=\frac{2}{d(d^2-1)(d^2-4)} =: c.
\end{align}
giving
($s=e+3t+2c$)~\cite{BrouwerBeenakker1996, PuchalaMiszczak2017}
\begin{equation}
    \label{eq:W}
    W(d) = \left[ \begin{array}{c|ccc|cc}
    e & t & t & t & c & c \\
    \hline
    t & e & c & c & t & t \\
    t & c & e & c & t & t \\
    t & c & c & e & t & t \\
    \hline
    c & t & t & t & e & c \\
    c & t & t & t & c & e \\
    \end{array} \right]\in\mathbb R^{6\times 6}.
\end{equation}
Therefore Eq.\eqref{eq:weingarten} can be reformulated as
\begin{equation}
 \boldsymbol y_{\rm patt}
 =S W(d)\boldsymbol x_{\rm full}
 \label{eq:local_full_map1}
\end{equation}
The two last columns of \(S W(d)\) associated with \((123)\) and \((132)\) coincide, hence they can be rewritten as one column multiplying the sum of last two entries of $\boldsymbol x_{\rm full}$.
Define the local cycle-merging map
\(C:\mathbb C^6\to\mathbb C^5\) by
\begin{align}
 C&(a_1,a_{12},a_{23},a_{13},a_{123},a_{132})^T
 \nonumber\\
 &=(a_1,a_{12},a_{23},a_{13},a_{123}+a_{132})^T.
 \label{eq:cycle_sum}
\end{align}
It retains the sum of the cycle coordinates and annihilates
their difference. For one party, those coordinates already
coincide, so this merging removes no information. Its information-losing consequence for distinct invariants
will appear in the two-party case.
Defining $M(d)$ as $SW(d)$ with the last column removed and $\boldsymbol z = C \boldsymbol x_{\rm full}$ one can rewrite the equation (\ref{eq:local_full_map1}) as
\begin{equation}
    \boldsymbol y_{\rm patt} = M(d) \boldsymbol z,
    \label{eq:local_full_map}
\end{equation}
where
\begin{equation}
 M(d)=
 \begin{pmatrix}
 e&t&t&t&c\\
 e+t&e+t&c+t&c+t&c+t\\
 e+t&c+t&e+t&c+t&c+t\\
 e+t&c+t&c+t&e+t&c+t\\
 s&s&s&s&s
 \end{pmatrix}.
 \label{eq:local_map}
\end{equation}

We label the retained coordinates in $\boldsymbol z$ by \(\Pi\), matching the
identity to \(D\), each transposition \((ab)\) to \(T_{ab}\),
and the cycle sum to \(E\). These labels match coordinates
with different meanings: \(E\) denotes a cycle sum in
\(\boldsymbol z\), whereas three equal outcomes in $\boldsymbol{y}_{\rm patt}$.

The map $M(d)$ is the local building block used
at every party in the multipartite construction. The pole at \(d=2\) in Eq.~\eqref{eq:Wg_values_e}-\eqref{eq:Wg_values_c} occurs because one of the five outcome patterns is absent for qubits~\cite{Scala2027}.

\subsection{Orbit reduction}

Relabelling the three identical copies does not change a
contraction. Indeed,
\(P_\eta P_\tau P_\eta^{-1}=P_{\eta\tau\eta^{-1}}\) and
\(P_\eta\rho^{\otimes3}P_\eta^{-1}=\rho^{\otimes3}\), so
\begin{equation}
 x(\eta\tau\eta^{-1})=x(\tau),
 \qquad \eta\in S_3.
 \label{eq:contraction_conjugation_invariance}
\end{equation}
Thus contractions are equal on each conjugation orbit.

The same relabelling leaves the probability product unchanged
and acts on equality patterns as
\begin{equation}
 \eta\cdot D=D,\qquad
 \eta\cdot E=E,\qquad
 \eta\cdot T_{ab}=T_{\eta(a)\eta(b)}.
 \label{eq:pattern_relabelling}
\end{equation}
There are therefore three pattern orbits: \(D\), the three
\(T\) labels, and \(E\). Their data coordinates are
\(y_D,y_T,y_E\), where \(y_T\) is the common value of the
three labelled \(T\) entries.

As contractions on an orbit are equal, we change the equality-pattern-labelled coordinates to orbit-labelled coordinates, summing in $M(d)$ the columns related to permutations of the same orbit. The column-merging is realised by the right multiplication with the matrix $R_1$:
\begin{equation}
 R_1=\begin{pmatrix}
 1&0&0\\ 0&1&0\\ 0&1&0\\ 0&1&0\\ 0&0&1
 \end{pmatrix},\quad
M(d) R_1 = \begin{pmatrix}
e&3t&c \\ e+t&s&c+t \\ e+t&s&c+t \\ e+t&s&c+t \\ s&3s&s 
 \end{pmatrix}.
 \label{eq:one_party_orbit_maps}
\end{equation}
The same applies to renumbering of outcomes in $\boldsymbol y_{\rm patt}$ -- these belonging to the same permutation orbit are equal and hence are realised by identical rows in $M(d) R_1$. We remove the duplicates by averaging over orbits:
\begin{equation}
L_1=\begin{pmatrix}
 1&0&0&0&0\\
 0&\frac13&\frac13&\frac13&0\\
 0&0&0&0&1
 \end{pmatrix},
 \quad L_1R_1=I_3.
 \label{eq:one_party_orbit_maps_L}
\end{equation}
Substituting
\(\boldsymbol z=R_1(q_0,q_1,2q_2)^T\) and applying from left \(L_1\) to
Eq.~\eqref{eq:local_map} gives the in the orbit space
\begin{align}
 &\begin{pmatrix}y_D\\y_T\\y_E\end{pmatrix}
 =
 H(d)
 \begin{pmatrix}q_0\\q_1\\2q_2\end{pmatrix},
 \\
 &H(d)=
 \begin{pmatrix}
 e&3t&c\\
 e+t&s&c+t\\
 s&3s&s
 \end{pmatrix}=L_1M(d)R_1.
 \label{eq:one_system}
\end{align}

The factor \(3\) collects the three identical transposition
coordinates, whereas \(2q_2\) retains the cycle-sum convention.
To reconstruct the invariant coordinates \(q_0,q_1,2q_2\), one only needs to
invert the \(3\times3\) one-party matrix \(H(d)\):
\begin{equation}
 H(d)^{-1}=
 \begin{pmatrix}
 d(d-1)(d-2)&3d(d-1)&d\\
 -d(d-1)(d-2)&d(d-1)(d-2)&d^2\\
 2d(d-1)(d-2)&-3d(d-1)^2&d^3+d
 \end{pmatrix}.
 \label{eq:one_inverse}
\end{equation}
The striking consequence, which we prove at the end of the paper, is that the
many-party reconstruction problem collapses to this same \(3\times3\)
inverse and scalar blocks. Thus the apparent growth of the full reconstruction matrix, studied in the proper basis, does
not introduce any new nontrivial matrix inversion.

For a normalized
state, \(q_0=1\); retaining this coordinate keeps the cubic
map homogeneous and makes its normalization explicit.
The matrix \(H(d)\) will also form the blocks of the
multipartite reconstruction.

Finally, representative correlators are converted to event
probabilities by counting outcome realizations. The local
multiplicities are
\begin{equation}
 m_D(d)=d(d-1)(d-2),\quad
 m_T(d)=d(d-1),\quad
 m_E(d)=d.
 \label{eq:multiplicity}
\end{equation}
There are three labelled \(T\) patterns, so normalization reads
\begin{equation}
 q_0=m_D(d)y_D+3m_T(d)y_T+m_E(d)y_E=1.
 \label{eq:normalization}
\end{equation}
These multiplicities convert measured pattern frequencies
into the representative coordinates required by the inverse
map. Appendix~\ref{app:frequencies} extends this conversion
to multipartite orbit events.
\section{Two parties}\label{sec:two_parties}

\subsection{Equality patterns and local cycle merging}

For two parties, independent local Haar integration gives the
tensor product of the one-party maps. The new ingredient is
the relative arrangement of the copy labels at \(A\) and \(B\).

Let \(\rho\) act on
\(\mathbb C^{d_A}\otimes\mathbb C^{d_B}\). Grouping the three
copies of each party together, define
\begin{equation}
 x(\tau_A,\tau_B)
 =
 \operatorname{Tr}\!\left[
 (P_{\tau_A}^{A}\otimes P_{\tau_B}^{B})\rho^{\otimes3}
 \right],
 \quad (\tau_A,\tau_B)\in S_3^2.
 \label{eq:contractions_two}
\end{equation}
The vector \(\boldsymbol x_{\rm full}^{(2)}\in\mathbb C^{36}\)
collects these labelled contractions. Each party also supplies
one equality-pattern label, giving
\(\boldsymbol y_{\rm patt}^{(2)}\in\mathbb R^{25}\).
Tensoring the one-party relation \(S W(d)=M(d)C\) yields
\begin{align}
 \boldsymbol y_{\rm patt}^{(2)}
 &=[S W(d_A)\otimes S W(d_B)]
   \boldsymbol x_{\rm full}^{(2)}
 \label{eq:two_local_system}\\
 &=[M(d_A)\otimes M(d_B)]\boldsymbol z^{(2)},
 \qquad
 \boldsymbol z^{(2)}
 =(C\otimes C)\boldsymbol x_{\rm full}^{(2)}.
 \nonumber
\end{align}
Thus the same local cycle-merging map acts independently
on the contraction label at each party.

A coordinate without an \(E\) label is unchanged by $C^{\otimes 2}$; for example,
\(z_{T_{12}T_{23}}^{(2)}=x((12),(23))\).
One \(E\) label sums two cycle orientations,
\begin{equation}
 z_{ET_{12}}^{(2)}
 =
 x((123),(12))+x((132),(12)),
\end{equation}
whereas two \(E\) labels sum all four orientation pairs:
\begin{align}
 z_{EE}^{(2)}
 &=x((123),(123))+x((123),(132))
 \nonumber\\
 &\quad+x((132),(123))+x((132),(132)).
 \label{eq:two_party_four_cycles}
\end{align}
The measurement map therefore retains this sum before any
orbit reduction is performed. The orbit classification below
identifies the physical invariants entering it.

\subsection{Orbit reduction}

The three copy positions label complete bipartite
preparations, so a relabelling must act simultaneously at
both parties. The resulting diagonal conjugation action is
\begin{equation}
 \eta\cdot(\tau_A,\tau_B)
 =
 (\eta\tau_A\eta^{-1},\eta\tau_B\eta^{-1}),
 \qquad \eta\in S_3.
 \label{eq:diagonal_conjugation_action}
\end{equation}
Invariance due to identical copies $\rho^{\otimes 3}$, as in the one-party case, gives
\(x(\eta\cdot(\tau_A,\tau_B))=x(\tau_A,\tau_B)\).
An orbit consists of the distinct pairs generated by these
six relabellings.
There are eleven contraction orbits. Table~\ref{tab:two_invariants}
lists their representative values containing the physical information -- the local unitary invariants of $\rho$.
\begin{table}[t]
\centering
\small
\begin{tabular}{lc}
\hline\hline
Contraction coordinate & Orbit size\\
\hline
\(x_0=x(1,1)=(\operatorname{Tr}\rho)^3\) & \(1\)\\
\(x_1=x(1,\tau)
 =(\operatorname{Tr}\rho)\operatorname{Tr}\rho_B^2\) & \(3\)\\
\(x_2=x(1,\kappa)=\operatorname{Tr}\rho_B^3\) & \(2\)\\
\(x_3=x(\tau,1)
 =(\operatorname{Tr}\rho)\operatorname{Tr}\rho_A^2\) & \(3\)\\
\(x_4=x(\tau,\tau')
 =\operatorname{Tr}[\rho(\rho_A\otimes\rho_B)]\) & \(6\)\\
\(x_5=x(\tau,\tau)
 =(\operatorname{Tr}\rho)\operatorname{Tr}\rho^2\) & \(3\)\\
\(x_6=x(\tau,\kappa)
 =\operatorname{Tr}[(\operatorname{Tr}_A\rho^2)\rho_B]\) & \(6\)\\
\(x_7=x(\kappa,1)=\operatorname{Tr}\rho_A^3\) & \(2\)\\
\(x_8=x(\kappa,\tau)
 =\operatorname{Tr}[(\operatorname{Tr}_B\rho^2)\rho_A]\) & \(6\)\\
\(x_9=x(\kappa,\kappa)=\operatorname{Tr}\rho^3\) & \(2\)\\
\(x_{10}=x(\kappa,\kappa^{-1})
 =\operatorname{Tr}[(\rho^{\Gamma_B})^3]\) & \(2\)\\
\hline\hline
\end{tabular}
\caption{
The eleven diagonal-conjugation orbits of bipartite
contractions~\cite{QST2025,scala2026finite}.
Here \(\tau=(12)\), \(\tau'=(23)\), \(\kappa=(123)\),
\(\rho_A=\operatorname{Tr}_B\rho\), and
\(\rho_B=\operatorname{Tr}_A\rho\).
The operation \(\Gamma_B\) is partial transposition on \(B\).
}
\label{tab:two_invariants}
\end{table}

In particular, equal cycle orientations give \(x_9\), while
opposite orientations give \(x_{10}\). Each class occurs
twice in Eq.~\eqref{eq:two_party_four_cycles}, so
\begin{equation}
 z_{EE}^{(2)}=2(x_9+x_{10}).
 \label{eq:cycle_pair_sum}
\end{equation}
Unlike the two cycle coordinates for one party, \(x_9\)
and \(x_{10}\) need not agree. The local cycle merging
therefore retains their sum but cannot separate their
difference.

On the data side, simultaneous relabelling acts on
equality-pattern pairs through
\begin{equation}
 \eta\cdot(\pi_A,\pi_B)
 =
 (\eta\cdot\pi_A,\eta\cdot\pi_B),
 \qquad (\pi_A,\pi_B)\in\Pi^2,
 \label{eq:diagonal_orbit_pattern}
\end{equation}

The eight local-type pairs other than \(TT\) each form
one orbit. The \(TT\) pairs split into two orbits according
to whether the equalities involve the same pair of shots,
\(T{=}T\), or different pairs, \(T{\ne}T\).
Table~\ref{tab:two_pattern_orbits} gives all ten orbits.
\begin{table}[t]
\centering
\small
\begin{tabular}{clc}
\hline\hline
Orbit & Equality pattern of $\Pi^2$ & Size\\
\hline
\(DD\) & \(\{(D,D)\}\) & \(1\)\\
\(DT\) & \(\{(D,T_{12}),(D,T_{23}),(D,T_{13})\}\) & \(3\)\\
\(DE\) & \(\{(D,E)\}\) & \(1\)\\
\(TD\) & \(\{(T_{12},D),(T_{23},D),(T_{13},D)\}\) & \(3\)\\
\(T{=}T\) & \(\{(T_{ab},T_{ab}):ab\in\{12,23,13\}\}\) & \(3\)\\
\(T{\ne}T\) & \(\{(T_{ab},T_{cd}):ab\ne cd\}\) & \(6\)\\
\(TE\) & \(\{(T_{12},E),(T_{23},E),(T_{13},E)\}\) & \(3\)\\
\(ED\) & \(\{(E,D)\}\) & \(1\)\\
\(ET\) & \(\{(E,T_{12}),(E,T_{23}),(E,T_{13})\}\) & \(3\)\\
\(EE\) & \(\{(E,E)\}\) & \(1\)\\
\hline\hline
\end{tabular}
\caption{
The ten orbits of labelled equality-pattern pairs.
The pair labels \(ab,cd\) range over \(\{12,23,13\}\).
Common relabelling preserves whether the parties select
the same or different pairs of shots.
}
\label{tab:two_pattern_orbits}
\end{table}
In the table order, the representative data values and
matching merged contractions are
\begin{align}
 \boldsymbol y_{\rm orb}^{(2)}
 &=(y_{DD},y_{DT},y_{DE},y_{TD},y_{T=T},y_{T\ne T},
 \nonumber\\
 &\hspace{2.5em}
 y_{TE},y_{ED},y_{ET},y_{EE})^T,
 \nonumber\\
 \boldsymbol z_{\rm orb}^{(2)}
 &=(x_0,x_1,2x_2,x_3,x_5,x_4,
 \nonumber\\
 &\hspace{2.5em}2x_6,2x_7,2x_8,2(x_9+x_{10}))^T.
 \label{eq:two_orbit_vectors}
\end{align}
The factors of two follow from \(C^{\otimes 2}\): each
single-\(E\) coordinate sums two equal contractions, while
the \(EE\) coordinate contains two copies of each cycle
orbit. 
The normalization convention for the data coordinates $\boldsymbol y_{\rm orb}^{(2)}$  is the same as
in the one-party case of Eq.~\eqref{eq:normalization}.

To express the map that relates data and invariants in these ten coordinates for both $\boldsymbol y_{\rm orb}^{(2)}$ and $\boldsymbol z_{\rm orb}^{(2)}$, let
\(\mathcal O_1,\ldots,\mathcal O_{10}\) be the pattern
orbits in Table~\ref{tab:two_pattern_orbits}. Generalize $R_1$ and $L_1$,
the embedding and extraction maps of Eqs.~\eqref{eq:one_party_orbit_maps} and \eqref{eq:one_party_orbit_maps_L}, by
\begin{equation}
 (R_2)_{q\mu}=\mathbf 1_{\{q\in\mathcal O_\mu\}},
 \qquad
 (L_2)_{\mu q}
 =\frac{\mathbf 1_{\{q\in\mathcal O_\mu\}}}
 {|\mathcal O_\mu|},
 \label{eq:two_orbit_maps}
\end{equation}
where \(q\in\Pi^2\) and \(\mu=1,\ldots,10\).
Here \(R_2:\mathbb R^{10}\to\mathbb R^{25}\) embeds the ten orbit
coordinates into the \(25\) labelled coordinates by assigning one if the pair $q$ belongs to the orbit $\mathcal{O}_\mu$. 
Conversely, \(L_2:\mathbb R^{25}\to\mathbb R^{10}\) extracts one coordinate per orbit by
averaging the labelled entries in that orbit. Hence \(L_2R_2=I_{10}\).

\begin{figure}[t]
\centering
\begin{tikzpicture}[
  box/.style={
    draw,
    rounded corners=2pt,
    align=center,
    text width=0.39\columnwidth,
    minimum height=1.05cm,
    inner sep=3pt,
    font=\footnotesize
  },
  arrow/.style={-{Latex[length=1.6mm]},thin},
  reduction/.style={
    font=\scriptsize,
    align=center,
    fill=white,
    inner sep=2pt
  }
]

\node[box] (x) {
  Full contractions\\
  $\boldsymbol x_{\rm full}^{(n)}$\\
  $6^n$ entries
};
\node[box,right=0.10\columnwidth of x] (yout) {
  Averaged outcome data\\
  $\boldsymbol y_{\rm out}^{(n)}$\\
  $\prod_k d_k^3$ entries
};

\node[box,below=1.05cm of x] (z) {
  Merged contractions\\
  $\boldsymbol z^{(n)}$\\
  $5^n$ entries
};
\node[box,below=1.05cm of yout] (ypatt) {
  Labelled pattern data\\
  $\boldsymbol y_{\rm patt}^{(n)}$\\
  $5^n$ entries
};

\node[box,below=1.05cm of z] (zorb) {
  Merged contraction orbits\\
  $\boldsymbol z_{\rm orb}^{(n)}$\\
  $N_n$ coordinates
};
\node[box,below=1.05cm of ypatt] (yorb) {
  Pattern-orbit data\\
  $\boldsymbol y_{\rm orb}^{(n)}$\\
  $N_n$ coordinates
};

\node[above=3mm of x,font=\footnotesize\bfseries]
  {State invariants};
\node[above=3mm of yout,font=\footnotesize\bfseries]
  {Measured data};

\draw[arrow] (x) -- (yout);
\draw[arrow] (z) -- node[above,font=\scriptsize] {$M_n$} (ypatt);
\draw[arrow] (zorb) --
  node[above,font=\scriptsize] {$M_{\rm orb}$} (yorb);

\draw[arrow] (x) -- node[reduction] {
  Local cycle merging\\
  $C^{\otimes n}$
} (z);
\draw[arrow] (yout) -- node[reduction] {
  Equality-pattern\\
  reduction
} (ypatt);

\draw[arrow] (z) -- node[reduction] {
  Orbit reduction\\
  $L_n$
} (zorb);
\draw[arrow] (ypatt) -- node[reduction] {
  Orbit reduction\\
  $L_n$
} (yorb);

\end{tikzpicture}
\caption{
From full labelled coordinates to the reconstruction map \(M_{\rm orb}\).
Horizontal arrows relate state invariants to measured averages through
Haar integration. Equality-pattern reduction removes repeated outcome
entries, whereas local cycle merging by $C^{\otimes n}$ adds the two cycle orientations (e.g. \eqref{eq:cycle_pair_sum})
and removes access to their differences. Simultaneous copy relabelling
then gives one coordinate per orbit in each column, without further
information loss. The resulting map \(M_{\rm orb}\) is reconstructed
blockwise by the theorem.
}
\label{fig:funnel}
\end{figure}
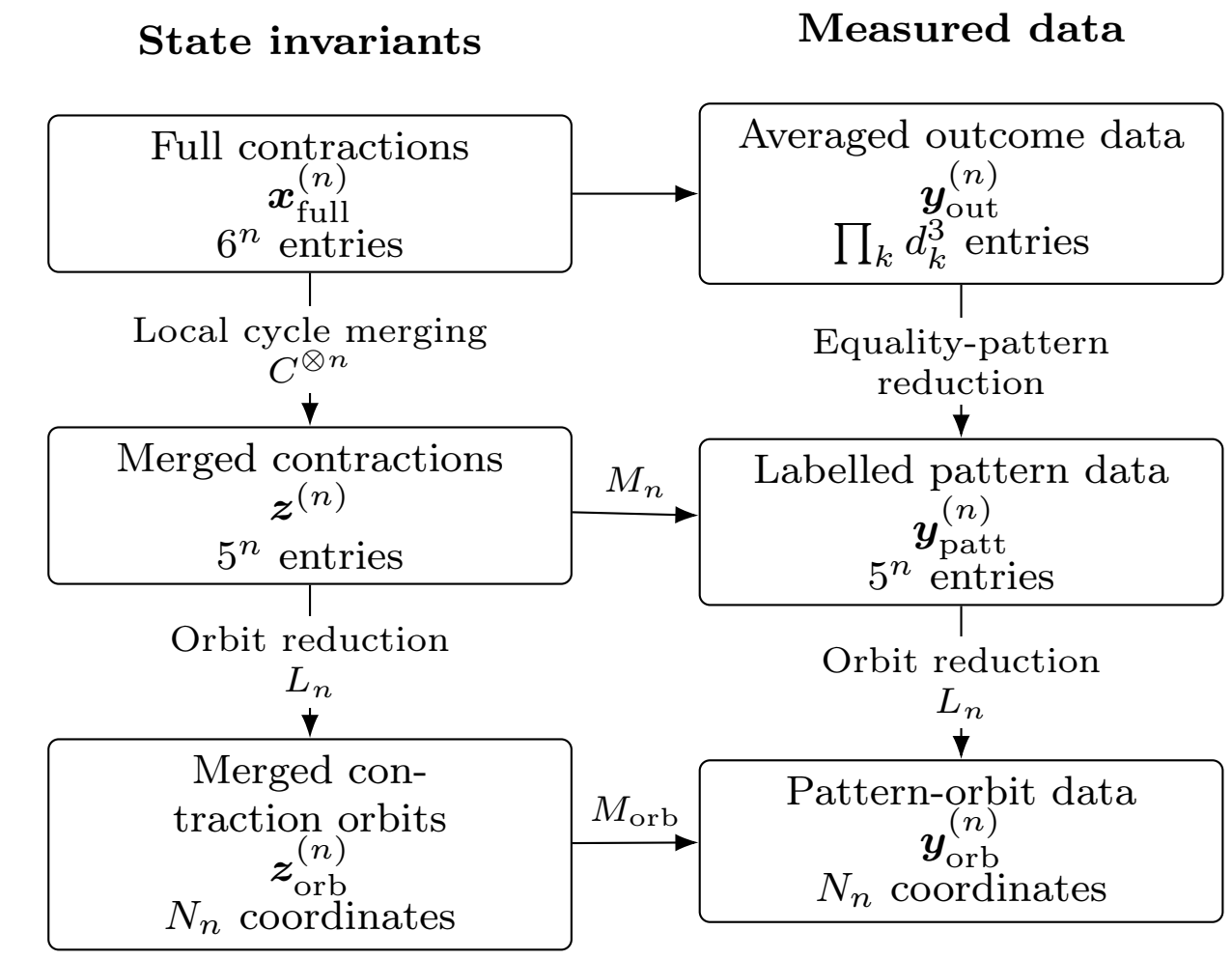

Both labelled vectors $\boldsymbol y_{\rm patt}^{(2)}$ and $\boldsymbol z^{(2)}$ are constant on the orbits, so
\begin{align}
 \boldsymbol z^{(2)}
 &=R_2\boldsymbol z_{\rm orb}^{(2)},
 \qquad
 \boldsymbol y_{\rm orb}^{(2)}
 =L_2\boldsymbol y_{\rm patt}^{(2)},\nonumber\\
 \boldsymbol z^{(2)}_{\rm orb}
 &=L_2\boldsymbol z^{(2)},
 \qquad
 \boldsymbol y_{\rm patt}^{(2)}
 =R_2\boldsymbol y_{\rm orb}^{(2)}\nonumber.
\end{align}
Substituting the first relation into
Eq.~\eqref{eq:two_local_system} and applying \(L_2\) gives
\begin{align}
 \boldsymbol y_{\rm orb}^{(2)}
 &=M_{\rm orb}(d_A,d_B)\boldsymbol z_{\rm orb}^{(2)},
 \nonumber\\
 M_{\rm orb}(d_A,d_B)
 &=L_2[M(d_A)\otimes M(d_B)]R_2.
 \label{eq:two_reduced_system}
\end{align}
This orbit reduction removes repeated coordinates without
further loss of the information retained by \(C\otimes C\).
Figure~\ref{fig:funnel} summarizes these steps in the
multipartite notation introduced in the next section and the solution for $n=2$ that expresses $\boldsymbol z^{(2)}_{\rm orb}$ from $\boldsymbol y^{(2)}_{\rm orb}$ will be obtained as an instance of the general theorem for arbitrary $n$ in Sec. \ref{sec:bipartite_case}.
\section{Arbitrary parties}\label{sec:general}

\subsection{Equality patterns and local cycle merging}

The two-party construction extends by applying the same local
maps $S\,W(d_i)$ at every party. A tuple
\(\boldsymbol\tau=(\tau_1,\ldots,\tau_n)\in S_3^n\)
specifies the contraction
\begin{equation}
 x_\rho(\boldsymbol\tau)
 =
 \operatorname{Tr}\!\left[
 \left(\bigotimes_{k=1}^nP_{\tau_k}^{(k)}\right)
 \rho^{\otimes3}\right],
 \qquad
 \boldsymbol x_{\rm full}^{(n)}\in\mathbb C^{6^n}.
 \label{eq:contractions_general}
\end{equation}

As before in Eq.~\eqref{eq:contractions_two}, tensor factors are grouped by party.
For arbitrary party number, the full contractions need not
be real. Since \(\rho^{\otimes3}\) is Hermitian and
\(P_\tau^\dagger=P_{\tau^{-1}}\), complex conjugation gives
\begin{equation}
 \overline{x_\rho(\tau_1,\ldots,\tau_n)}
 =
 x_\rho(\tau_1^{-1},\ldots,\tau_n^{-1}).
 \label{eq:complex_conjugation}
\end{equation}
For one or two parties, the inverse tuple belongs to the same
diagonal-conjugation orbit, so each contraction is real.
For three or more parties, this need not hold.

The merged vector
\(\boldsymbol z^{(n)}=C^{\otimes n}
\boldsymbol x_{\rm full}^{(n)}\) nevertheless remains real.
To see this, write the operators associated with its local
labels as
\begin{equation}
    O_D=I,\qquad
 O_{T_{ab}}=P_{(ab)},\qquad
 O_E=P_{(123)}+P_{(132)}.
 \label{eq:retained_op}
\end{equation}
Each component is then
\begin{equation}
 z_{\boldsymbol\pi}^{(n)}
 =
 \operatorname{Tr}\!\left[
 \left(\bigotimes_{k=1}^nO_{\pi_k}^{(k)}\right)
 \rho^{\otimes3}\right],
 \qquad \boldsymbol\pi\in\Pi^n.
 \label{eq:merged_real}
\end{equation}
Both operators inside the trace are Hermitian, so cyclicity
of the trace makes this expression real.

Denote the one-party coordinate space by
\(A=\operatorname{span}_{\mathbb R}
\{e_\pi:\pi\in\Pi\}\cong\mathbb R^5\), with its standard
orthonormal basis. Tensoring the local relation
\(S W(d)=M(d)C\) gives
\begin{equation}
 \boldsymbol y_{\rm patt}^{(n)}=M_n\boldsymbol z^{(n)},
 \qquad
 M_n=\bigotimes_{k=1}^nM(d_k):
 A^{\otimes n}\to A^{\otimes n}.
 \label{eq:general_pattern_system}
\end{equation}
Although the map $M_n$ factorizes, the input  and data 
vectors, respectively $\boldsymbol z^{(n)}$ and $\boldsymbol y_{\rm patt}^{(n)}$, generally do not: both retain correlations among
the parties.

\subsection{Orbit reduction}

As for two parties, one common \(\eta\in S_3\) relabels
the three preparations. The local pattern action
in Eq.~\eqref{eq:pattern_relabelling} defines the matrices
\begin{equation}
 \gamma(\eta)e_\pi=e_{\eta\cdot\pi},
 \qquad
 \Gamma_n(\eta)=\gamma(\eta)^{\otimes n},
 \label{eq:gamma}
\end{equation}
acting on \(A\) and \(A^{\otimes n}\), respectively.
The retained operators in Eq.~\eqref{eq:retained_op} obey
\(P_\eta O_\pi P_\eta^{-1}=O_{\eta\cdot\pi}\)
defining the same action both for the pattern equality of experimental data and the invariant coordinates. Indeed, the action fixes \(I\), sends
\(P_{(ab)}\) to \(P_{(\eta(a)\eta(b))}\), and permutes the two
three-cycles while preserving their sum.

Relabelling the three probability factors leaves their product
unchanged. Likewise, relabelling identical global copies leaves
\(\rho^{\otimes3}\) unchanged. Thus
\begin{equation}
 \Gamma_n(\eta)\boldsymbol y_{\rm patt}^{(n)}
 =\boldsymbol y_{\rm patt}^{(n)},
 \qquad
 \Gamma_n(\eta)\boldsymbol z^{(n)}
 =\boldsymbol z^{(n)}.
 \label{eq:fixed_vectors}
\end{equation}
Both labelled vectors therefore lie in the invariant subspace
\((A^{\otimes n})^{S_3}\): their entries are constant on each
simultaneous-relabelling orbit in \(\Pi^n\). One coordinate per
orbit suffices. Denote the number of orbits by \(N_n\); its count
is derived in Appendix~\ref{app:characters}.

Extend the two-party embedding and extraction maps by setting,
for each orbit \(o\subset\Pi^n\),
\begin{equation}
 (R_n)_{\boldsymbol\pi,o}
 =\mathbf1_{\{\boldsymbol\pi\in o\}},
 \qquad
 (L_n)_{o,\boldsymbol\pi}
 =\frac{\mathbf1_{\{\boldsymbol\pi\in o\}}}{|o|}.
 \label{eq:orbit_maps}
\end{equation}
Here \(R_n:\mathbb R^{N_n}\to A^{\otimes n}\) embeds the \(N_n\) orbit
coordinates into the full labelled space by assigning one to
all labelled pattern tuples in the same diagonal orbit. Conversely,
\(L_n:A^{\otimes n}\to\mathbb R^{N_n}\) extracts one coordinate per orbit by
averaging over its labelled members. Consequently,
$L_nR_n=I_{N_n}$. Define
\begin{equation}
 \boldsymbol y_{\rm orb}^{(n)}
 =L_n\boldsymbol y_{\rm patt}^{(n)},
 \qquad
 \boldsymbol z_{\rm orb}^{(n)}
 =L_n\boldsymbol z^{(n)}.
 \label{eq:orbLpatt}
\end{equation}
Since all entries within an orbit agree, repeating each extracted
value recovers the original labelled vectors:
\begin{equation}
 \boldsymbol y_{\rm patt}^{(n)}
 =R_n\boldsymbol y_{\rm orb}^{(n)},
 \qquad
 \boldsymbol z^{(n)}
 =R_n\boldsymbol z_{\rm orb}^{(n)}.
 \label{eq:pattRorb}
\end{equation}
Thus these coordinate reductions are reversible on the
invariant subspace.

Substituting \(\boldsymbol z^{(n)}=R_n\boldsymbol z_{\rm orb}^{(n)}\)
into Eq.~\eqref{eq:general_pattern_system} and applying \(L_n\)
gives
\begin{equation}
 \begin{aligned}
 \boldsymbol y_{\rm orb}^{(n)}
 &=L_nM_nR_n\boldsymbol z_{\rm orb}^{(n)}
 =M_{\rm orb}(\boldsymbol d)\boldsymbol z_{\rm orb}^{(n)},
 \end{aligned}
 \label{eq:orbit_system}
\end{equation}
where \(\boldsymbol d=(d_1,\ldots,d_n)\).
The dimension sum in
Eq.~\eqref{app:eq:dimension_sum} gives \(N_n\sim5^n/6\).
Storing a dense \(N_n\times N_n\) matrix requires
\(O(N_n^2)\) memory, and conventional elimination costs
\(O(N_n^3)\) arithmetic operations; both grow exponentially
with party number. A numerical inverse would recover the coordinates without revealing which
invariant combinations are actually independent. We therefore decompose the
map into symmetry-resolved sectors, where every nontrivial matrix inversion
reduces to the same \(3\times3\) one-party blocks, together with scalar
weights.
\section{Reconstruction sectors}\label{sec:sectors}

The orbit reduction identifies the independent coordinates of the data and
invariant vectors. In this section, we introduce a second coordinate system
that makes the reconstruction map easier to analyze and invert. 
\subsection{Local invariant and zero-sum subspaces}

Copy relabelling fixes \(D\) and \(E\), but permutes the three \(T\)
labels. We separate their common component from two orthogonal
zero-sum components by choosing
\begin{align}
 v_T&=e_{T_{12}}+e_{T_{23}}+e_{T_{13}},\nonumber\\
 \xi_1&=e_{T_{12}}-e_{T_{23}},\qquad
 \xi_2=e_{T_{12}}+e_{T_{23}}-2e_{T_{13}}.
\end{align}
In the ordered basis
\((e_D,e_{T_{12}},e_{T_{23}},e_{T_{13}},e_E)\), the
change-of-basis matrix is
\begin{equation}
 F=(e_D,v_T,e_E,\xi_1,\xi_2)
 =
 \begin{pmatrix}
 1&0&0&0&0\\
 0&1&0&1&1\\
 0&1&0&-1&1\\
 0&1&0&0&-2\\
 0&0&1&0&0
 \end{pmatrix}.
 \label{eq:local_basis_change}
\end{equation}
Its columns form an orthogonal basis, without normalizing the
\(T\) vectors:
\(F^TF=\operatorname{diag}(1,3,1,2,6)\).
For
\(\boldsymbol a=(a_D,a_{12},a_{23},a_{13},a_E)^T\),
the new coordinates are
\[
 F^{-1}\boldsymbol a=
 \begin{pmatrix}
 a_D\\
 (a_{12}+a_{23}+a_{13})/3\\
 a_E\\
 (a_{12}-a_{23})/2\\
 (a_{12}+a_{23}-2a_{13})/6
 \end{pmatrix}.
\]
The first three columns of \(F\) are \(R_1\), and the first
three rows of \(F^{-1}\) are \(L_1\).

In this basis, relabelling leaves the first three coordinates
unchanged. The remaining two transform among themselves:
\begin{equation}
 F^{-1}\gamma(\eta)F
 =I_3 \oplus \gamma_W(\eta),
 \label{eq:local_relabelling_blocks}
\end{equation}
where, for two generators of \(S_3\),
\[
 \gamma_W((12))
 =\frac12\begin{pmatrix}1&3\\1&-1\end{pmatrix},
 \qquad
 \gamma_W((23))
 =\frac12\begin{pmatrix}1&-3\\-1&-1\end{pmatrix}.
\]
If a vector \([\alpha,\beta]^T \in W\) were unchanged by both
generators, the two invariance equations would give \(\alpha=\beta=0\).
The vectors invariant under every relabelling are therefore
precisely those whose last two coordinates vanish:
 $A^{S_3}=V=\operatorname{span}\{e_D,v_T,e_E\}$,
 $W=\operatorname{span}\{\xi_1,\xi_2\}$, so that $A=V\oplus W$.
Relabelling preserves \(W\) as a subspace, although \(W\)
contains no nonzero vector invariant under every relabelling.

As \([M(d),\gamma(\eta)]=0\), then \(M(d)\) does not mix \(V\) and \(W\), hence it is block-diagonal with respect to this basis. Its \(2\times2\)
block commuting with both displayed matrices
\(\gamma_W((12))\) and \(\gamma_W((23))\) must be scalar;
this is the instance of Schur's lemma used here. The change
of basis gives
\begin{equation}
 F^{-1}M(d)F
 =
 H(d)\oplus\lambda(d)I_2,
 \label{eq:local_action}
\end{equation}
with $\lambda(d)=e(d)-c(d)=\frac{1}{d(d^2-1)}$.
The first block is the one-party map
\(H(d)=L_1M(d)R_1\). The second says that \(M(d)\)
multiplies every zero-sum vector by the same scalar.
Although \(W\) alone has no invariant vector, tensor products
of \(W\) factors can have invariant combinations, as the next
subsection shows.

\subsection{Invariant tensor sectors}
The orbit map still couples many coordinates. We now use the local
decomposition to identify subspaces on which the measurement map acts
independently; their invariant parts will become the diagonal blocks in the theorem.

The $n$-party orbit space $A^{\otimes n}$ splits into a direct sum of tensor sectors:
\begin{equation}
 A^{\otimes n}
 =
 (V \oplus W)^{\otimes n}
 =
 \bigoplus_{\boldsymbol X\in\{V,W\}^n}
 \mathcal H_{\boldsymbol X},
 \label{eq:all_sectors}
\end{equation}
where \(\mathcal H_{\boldsymbol X}=\bigotimes_{i=1}^nX_i\).
To select the component in each sector, use the local orthogonal projectors.
In the ordered basis
\((D,T_{12},T_{23},T_{13},E)\), they are
\begin{equation*}
 \begin{aligned}
P_V&=e_De_D^T+\frac13v_Tv_T^T+e_Ee_E^T=1\oplus\frac13
 \begin{pmatrix}1&1&1\\1&1&1\\1&1&1\end{pmatrix}\oplus1,\\
 P_W&=I_5-P_V=0_1\oplus \frac13
 \begin{pmatrix}
 2&-1&-1\\
 -1&2&-1\\
 -1&-1&2\\
 \end{pmatrix}\oplus 0_1.
 \end{aligned}
\end{equation*}
The factor \(1/3\) accounts for \(\|v_T\|^2=3\).
Thus \(P_V=R_1L_1\) keeps the \(D,E\) entries and replaces the three
\(T\) entries by their average; \(P_W\) keeps their zero-sum
deviations. For a choice \(\boldsymbol X\), the tensor projector
\(P_{\boldsymbol X}=\bigotimes_{i=1}^nP_{X_i}\) 
projects onto
\(\mathcal H_{\boldsymbol X}\). The local relations
\(P_VP_W=0\) and \(P_V+P_W=I_5\) give
\begin{equation*}
 P_{\boldsymbol X}P_{\boldsymbol X'}
 =\delta_{\boldsymbol X\boldsymbol X'}P_{\boldsymbol X}
 ,\qquad\sum_{\boldsymbol X\in\{V,W\}^n}P_{\boldsymbol X}
 =I_{5^n}.
\end{equation*}

Both local maps \(\gamma(\eta)\) and \(M(d)\) preserve \(V\)
and \(W\), by the 
Eq.~\eqref{eq:local_action}.
They therefore commute with \(P_V,P_W\). 
The commutation
property extends to each sector:
\begin{equation*}
 M_nP_{\boldsymbol X}
 =\bigotimes_{i=1}^n[M(d_i)P_{X_i}]
 =\bigotimes_{i=1}^n[P_{X_i}M(d_i)]
 =P_{\boldsymbol X}M_n,
\end{equation*}
and hence $M_n$ factorises by sectors (has block-diagonal form with blocks supported on $\mathcal H_{\boldsymbol X}$), as $\Gamma(\eta)$ do.

Preserving a sector does not mean that every vector in it is
unchanged by relabelling. On the other hand, the physical data and merged-contraction
vectors $\boldsymbol y_{\rm orb}$ and $\boldsymbol z_{\rm orb}$ are invariant by Eq.~\eqref{eq:fixed_vectors}, and their
sector components inherit this property.
Therefore, their sector components live in
\(\mathcal H_{\boldsymbol X}^{S_3}\) -- 
the subspace of vectors in the sector
invariant with respect to common relabelling. 
The group-average Reynolds projector
onto the full invariant space 
$\left( A^{\otimes n}\right)^{S_3} = \bigoplus_{\boldsymbol X} \mathcal H_{\boldsymbol X}^{S_3}$ is
\begin{equation}
 P_{\rm inv}^{(n)}
 =\frac16\sum_{\eta\in S_3}\Gamma_n(\eta)
 =R_nL_n.
 \label{eq:orbit_projector}
\end{equation}
Since
\(P_{\boldsymbol X}\) commutes with every \(\Gamma_n(\eta)\),
the product \(P_{\rm inv}^{(n)}P_{\boldsymbol X}\) projects onto
\(\mathcal H_{\boldsymbol X}^{S_3}\). 
$\Tr P^{(n)}_{\rm inv}$ tells us how
many independent reconstruction coordinates that sector supplies.
As \(M(d)\) commutes with
\(\gamma(\eta)\): the latter is the identity on \(V\), while
\(M(d)\) is scalar on \(W\). Tensoring the commutation relation gives
\(M_n\Gamma_n(\eta)=\Gamma_n(\eta)M_n\). Thus \(M_n\)
preserves the invariant subspace. In particular, its action on
the orbit embedding satisfies
\begin{equation}
 M_nR_n=R_nL_nM_nR_n=R_nM_{\rm orb}.
 \label{eq:orbit_intertwining}
\end{equation}
This identifies the map $M_{\rm orb}(\boldsymbol d)$ on the space of orbits as the restriction of
the map $M_n(\boldsymbol d)$ on the equality-pattern space to the invariant coordinates.
To count explicitly the dimension of the invariant subspace \(\mathcal{H}^{S_3}\),
define
\[
 I_V(\boldsymbol X)=\{i:X_i=V\},\qquad
 I_W(\boldsymbol X)=\{i:X_i=W\},
\]
and let \(m=|I_V(\boldsymbol X)|\), \(k=|I_W(\boldsymbol X)|\).
After grouping the \(V\) factors first, pointwise invariance
of \(V\) gives
\begin{equation}
 \mathcal H_{\boldsymbol X}^{S_3}
 \simeq
 V^{\otimes m}\otimes(W^{\otimes k})^{S_3}.
\end{equation}
Hence only the invariant subspace of $W^{\otimes k}$ is left to be identified.
The dimension of the invariant subspace, derived in
Appendix~\ref{app:characters}, is
\begin{equation}
 \begin{aligned}
 w_k&=\dim(W^{\otimes k})^{S_3}
 =\frac{2^k+3\delta_{k0}+2(-1)^k}{6},\\
 \dim\mathcal H_{\boldsymbol X}^{S_3}&=3^m w_k.
 \end{aligned}
 \label{eq:sector_dimensions}
\end{equation}
Here \(k=0\) denotes the one-dimensional empty tensor product.
The first values are \(w_0=1\), \(w_1=0\), and \(w_2=w_3=1\).
The vanishing value \(w_1=0\) means that a sector with just one
zero-sum factor carries no invariant component of the measured data.
Therefore the sectors contributing to the invariant subspace are
\begin{equation}
\mathscr S_n
 =
 \{\boldsymbol X\in\{V,W\}^n:
 |I_W(\boldsymbol X)|\ne1\}.
 \label{eq:Sn}
\end{equation}
For two parties, \(VV\) contributes nine coordinates and \(WW\)
contributes one; \(VW\) and \(WV\) contribute none; constituting the 10-dimensional orbit space.
For three parties, \(VVV\), the three placements of \(WWV\),
and \(WWW\) give \(27+3\times3+1=37\) coordinates.
More generally, the dimension sum
\begin{equation}
    \sum_{k=0}^n\binom nk3^{n-k}w_k=N_n \,,
\end{equation} 
agrees with the orbit
count derived in Appendix~\ref{app:characters}. Thus the
invariant sectors supply a complete coordinate basis for the
data already identified by orbit reduction.

\subsection{Sector coordinates}

The preceding subsection identifies the invariant subspace within each
tensor sector. To write the restriction of $M_{ \rm orb}(\boldsymbol d)$ to these subspaces, we
need a basis for their coordinates and a way to convert back to the
orbit coordinates defined by the measured data.

For each \(k\ne1\), choose a matrix
\(J_k:\mathbb R^{w_k}\to A^{\otimes k}\) whose independent
columns span \((W^{\otimes k})^{S_3}\), and set \(J_0=[1]\).
One explicit construction is to select \(w_k\) independent
columns of the projector
\begin{equation}
 P_{\rm inv}^{(k)}P_W^{\otimes k},
 \qquad
 P_{\rm inv}^{(k)}
 =\frac16\sum_{\eta\in S_3}\gamma(\eta)^{\otimes k}.
 \label{eq:fixed_W_projector}
\end{equation}
Its image is exactly
\(\operatorname{im}(P_{\rm inv}^{(k)}P_W^{\otimes k})
=(W^{\otimes k})^{S_3}\). The factor \(P_{\rm inv}^{(k)}\)
averages over the same relabelling at all \(k\) positions,
while \(P_W^{\otimes k}\) restricts each factor to \(W\).
The two projectors
commute, so their product is a projector onto an intersection of $\left(A^{\otimes n}\right)^{S_3}$ and $W^{\otimes k}$ and hence has the rank \(w_k\). 
As no non-zero invariant vector exists in $W$, a single $W$ factor kills the invariant subspace of a tensor sector.

At each \(V\) position we pick one of the three columns of \(R_1\);
for the $k$ \(W\) positions jointly, we use one of the \(w_k\) columns of
\(J_k\). Taking all tensor products of these basis vectors and
restoring the original party order gives a basis of
\(\mathcal H_{\boldsymbol X}^{S_3}\) with \(3^m w_k\) columns.
The embedding of \(\mathcal H_{\boldsymbol X}^{S_3}\) is:
\begin{equation}
 R_{\boldsymbol X}^{\rm sec}
 =
 \Sigma_{\boldsymbol X}^{-1}
 (R_1^{\otimes m}\otimes J_k)
 \in\mathbb R^{5^n\times3^m w_k}.
 \label{eq:sector_basis}
\end{equation}
Its columns form a basis of
\(\mathcal H_{\boldsymbol X}^{S_3}\).
Indeed, both \(R_1^{\otimes m}\) and \(J_k\) have independent columns,
and \(\Sigma_{\boldsymbol X}\) only permutes tensor factors.
The inverse permutation restores the original party order;
the sector coordinates retain the column order of
\(R_1^{\otimes m}\otimes J_k\).
This product convention determines the tensor form of
the blocks in the theorem.

Each \(R_{\boldsymbol X}^{\rm sec}\) embeds coordinates from one
sector. To describe the embedding of all invariant spaces of subsequent sectors, one performs the horizontal concatenation of the sector embeddings: $R_n^{\rm sec}=\bigsqcup_{\boldsymbol X \in \mathscr S_n} R_{\boldsymbol{X}}^{\rm sec} \in\mathbb R^{5^n\times N_n}$.

\begin{figure}[t]
\centering
\begin{tikzpicture}[
  box/.style={draw,rounded corners=2pt,align=center,
    text width=0.30\columnwidth,minimum height=0.93cm,
    inner sep=3pt,font=\footnotesize},
  arrow/.style={-{Latex[length=1.mm]},thin},
  maplabel/.style={font=\small,fill=white,inner sep=1pt}
]
\node[box] (zorb) {Retained invariant orbits\\
  $\boldsymbol z_{\rm orb}^{(n)}$};
\node[box,right=0.20\columnwidth of zorb] (yorb)
  {Measured pattern orbits\\$\boldsymbol y_{\rm orb}^{(n)}$};
\node[box,below=.7cm of zorb] (zsec)
  {Invariant sectors\\$\boldsymbol z_{\rm sec}^{(n)}$};
\node[box,below=.7cm of yorb] (ysec)
  {Data sectors\\$\boldsymbol y_{\rm sec}^{(n)}$};
\draw[arrow] (zorb) -- node[maplabel,above] {$M_{\rm orb}$} (yorb);
\draw[arrow] (zorb) -- node[maplabel,left] {$Q_n^{-1}$} (zsec);
\draw[arrow] (yorb) -- node[maplabel,right] {$Q_n^{-1}$} (ysec);
\draw[arrow] (zsec) -- node[maplabel,above]
  {$\displaystyle\bigoplus_{\boldsymbol X\in\mathscr S_n}B_{\boldsymbol X}$}
  (ysec);
\end{tikzpicture}
\caption{From the orbit coordinates at the bottom of
Fig.~\ref{fig:funnel} to independent reconstruction sectors.
The vertical arrows are reversible changes of coordinates, with
the same matrix for the retained invariants and measured data.
}
\label{fig:sector_change}
\end{figure}
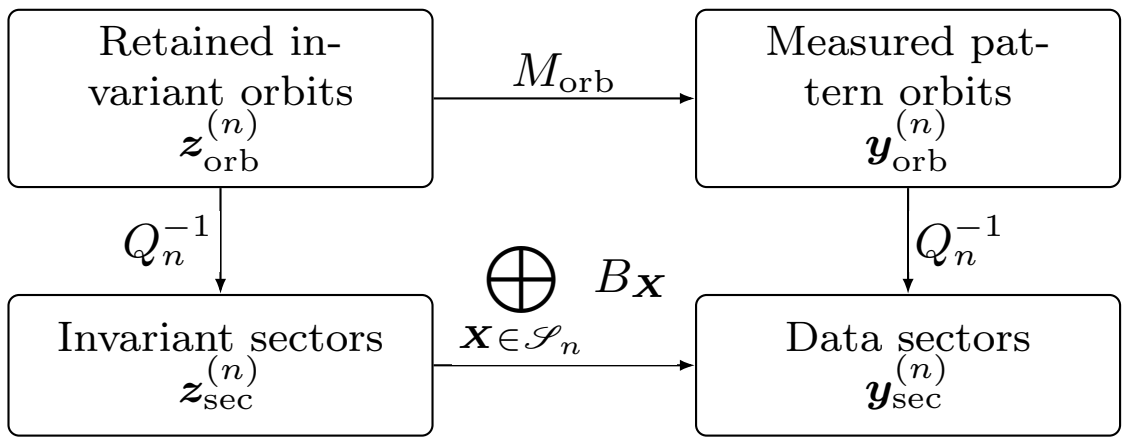

Because orbits do not change under the action of the group, the orbit space is a subspace of $\left(A^{\otimes n}\right)^{S_3}$. While the dimension of the orbit space and the invariant space agree, it is the same subspace.
The two embeddings $R_n$ and $R_n^{\rm sec}$ assign different coordinates
to the same vector: \(R_n\) uses orbit basis, whereas
\(R_n^{\rm sec}\) uses the sector bases. 
We need the matrix that converts the latter coordinates into the former.
Applying the orbit
extraction map \(L_n\) to each column of \(R_n^{\rm sec}\) gives
\begin{equation}
 Q_n=L_nR_n^{\rm sec},\qquad
 R_n^{\rm sec}=R_nQ_n.
 \label{eq:Q}
\end{equation}
The second equality holds because \(R_nL_n\) acts as the
identity on the invariant subspace, spanned by the columns of
\(R_n^{\rm sec}\). Thus \(Q_n\) changes coordinates 
of a fixed vector in $\left(A^{\otimes n}\right)^{S_3}$
from sector coordinates to orbit coordinates:
\[
 \boldsymbol y_{\rm orb}^{(n)}
 =Q_n\boldsymbol y_{\rm sec}^{(n)},\qquad
 \boldsymbol z_{\rm orb}^{(n)}
 =Q_n\boldsymbol z_{\rm sec}^{(n)}.
\]
Vice versa, define
\begin{equation}
 L_{\boldsymbol X}^{\rm sec}
 =\bigl[(R_{\boldsymbol X}^{\rm sec})^T
 R_{\boldsymbol X}^{\rm sec}\bigr]^{-1}
 (R_{\boldsymbol X}^{\rm sec})^T.
 \label{eq:sector_dual}
\end{equation}
The concatenation of these left inverses vertically in the same sector
order forms \(L_n^{\rm sec}\). Orthogonality between sectors
then gives \(L_n^{\rm sec}R_n^{\rm sec}=I_{N_n}\), while
\(R_n^{\rm sec}L_n^{\rm sec}=P_{\rm inv}^{(n)}\). Consequently,
 $Q_n^{-1}=L_n^{\rm sec}R_n$,
and \(Q_n^{-1}Q_n
=L_n^{\rm sec}R_nL_nR_n^{\rm sec}=I_{N_n}\)
by Eq.~\eqref{eq:orbit_projector}. 
The inverse \(Q_n^{-1}\) converts measured orbit data into the
sector coordinates on the direct sum of
\(H(d_i)\) and
\(\lambda(d_i)\).
This is the change of basis used in the main theorem below as represented in Fig.~\ref{fig:sector_change}, which
completes the path begun in
Fig.~\ref{fig:funnel}.

\section{Structural theorem}\label{sec:theorem} 
The theorem shows how to obtain the physical information in $\boldsymbol z^{(n)}=C^{\otimes n}\boldsymbol x_{\rm full}^{(n)}$ from 
the averaged data $\boldsymbol y_{\rm out}$ of Eq.~\eqref{eq:haar_data}
analytically.
\begin{theorem}
\label{thm:main}
Let \(n\ge1\), and use the orbit map
\(M_{\rm orb}(\boldsymbol d)\) and sector coordinate change \(Q_n\)
defined in Eqs.~\eqref{eq:orbit_system} and \eqref{eq:Q}.
For each contributing sector \(\boldsymbol X\in\mathscr S_n\), put $k=|I_W(\boldsymbol X)|$ and 
\[
 \Lambda_{\boldsymbol X}
 = \left\{
 \begin{array}{ll}
 1 & |I_W(\boldsymbol X)|= 0,\\
 \prod_{j\in I_W(\boldsymbol X)}\lambda(d_j) & |I_W(\boldsymbol X)|>1\\
 \end{array}
 \right.
\]
In the product basis of Eq.~\eqref{eq:sector_basis},
\begin{equation}
 \begin{aligned}
 &Q_n^{-1}M_{\rm orb}(\boldsymbol d)Q_n
 =\bigoplus_{\boldsymbol X\in\mathscr S_n}B_{\boldsymbol X},\\
 &\quad B_{\boldsymbol X}
 =\Lambda_{\boldsymbol X}
 \Sigma_{\boldsymbol X}^{-1}
 \left[
 \left(\bigotimes_{i\in I_V(\boldsymbol X)}H(d_i)\right)
 \otimes I_{w_k}
 \right]
 \Sigma_{\boldsymbol X}
 .
 \end{aligned}
 \label{eq:block_theorem}
\end{equation}
The block \(B_{\boldsymbol X}\) has dimension
\(3^{|I_V(\boldsymbol X)|}w_k\). 
Because each \(H(d_i)\) is invertible and each
\(\lambda(d_j)\) is nonzero, \(M_{\rm orb}(\boldsymbol d)\)
is invertible.
\end{theorem}
\begin{proof}
Fix a contributing sector \(\boldsymbol X\in\mathscr S_n\),
and retain its factor sets \(I_V,I_W\) and their sizes \(m,k\)
from the sector construction. The operator
\(\Sigma_{\boldsymbol X}\) in Eq.~\eqref{eq:sector_basis}
groups its \(V\) factors before its \(W\) factors. Hence
\begin{equation}
 \Sigma_{\boldsymbol X}M_n\Sigma_{\boldsymbol X}^{-1}
 =
 \left(\bigotimes_{i\in I_V}M(d_i)\right)
 \otimes
 \left(\bigotimes_{j\in I_W}M(d_j)\right).
 \label{eq:proof_factor_reordering}
\end{equation}
This reordering keeps each dimension \(d_i\) attached to its
party; the dimensions need not be equal.
From the two local actions in Eq.~\eqref{eq:local_action}
then
\begin{align}
 \left(\bigotimes_{i\in I_V}M(d_i)\right)R_1^{\otimes m}
 &=R_1^{\otimes m}
   \left(\bigotimes_{i\in I_V}H(d_i)\right),
 \nonumber\\
 \left(\bigotimes_{j\in I_W}M(d_j)\right)J_k
 &=J_k\Lambda_{\boldsymbol X}I_{w_k}.
 \label{eq:proof_factor_actions}
\end{align}
The operator before \(J_k\) acts as the same scalar on all
of \(W^{\otimes k}\), and therefore on the columns of \(J_k\).
Substituting Eq.~\eqref{eq:proof_factor_actions} into the
sector embedding \eqref{eq:sector_basis}, and using
Eq.~\eqref{eq:proof_factor_reordering}, gives
\begin{equation}
 M_nR_{\boldsymbol X}^{\rm sec}
 =R_{\boldsymbol X}^{\rm sec}B_{\boldsymbol X},
 \label{eq:proof_sector_intertwining}
\end{equation}
with \(B_{\boldsymbol X}\) as stated in
Eq.~\eqref{eq:block_theorem}. 
Concatenating Eq.~\eqref{eq:proof_sector_intertwining}
over the contributing sectors yields
\begin{equation}
 M_nR_n^{\rm sec}
 =R_n^{\rm sec}
 \left(\bigoplus_{\boldsymbol X\in\mathscr S_n}
 B_{\boldsymbol X}\right).
 \label{eq:proof_sector_sum}
\end{equation}
The relation \eqref{eq:orbit_intertwining}
and the basis relation \eqref{eq:Q} also give
\begin{equation}
 M_nR_n^{\rm sec}
 =R_nM_{\rm orb}Q_n
 =R_n^{\rm sec}(Q_n^{-1}M_{\rm orb}Q_n).
 \label{eq:proof_orbit_sum}
\end{equation}
Applying the left inverse \(L_n^{\rm sec}\) from
Eq.~\eqref{eq:sector_dual} to
Eqs.~\eqref{eq:proof_sector_sum} and
\eqref{eq:proof_orbit_sum} proves the block identity
\eqref{eq:block_theorem}. Finally, every \(H(d_i)\) has
the inverse in Eq.~\eqref{eq:one_inverse}, and each
\(\lambda(d_j)\) in Eq.~\eqref{eq:local_action} is nonzero.
Thus every block is invertible. The basis change \(Q_n\)
then makes \(M_{\rm orb}\) invertible as well.
\end{proof}

Thus each sector satisfies its own reconstruction system,
\[
 \boldsymbol y_{{\rm sec},\boldsymbol X}^{(n)}
 =
 B_{\boldsymbol X}
 \boldsymbol z_{{\rm sec},\boldsymbol X}^{(n)}.
\]
Its inverse requires only the known one-party inverse matrices $H(d_i)^{-1}$
and a scalar rescaling $\lambda(d_i)^{-1}$:
\begin{equation}
 \begin{aligned}
 B_{\boldsymbol X}^{-1}
 &=\Lambda_{\boldsymbol X}^{-1}
 \left[
 \left(\bigotimes_{i\in I_V(\boldsymbol X)}H(d_i)^{-1}\right)
 \otimes I_{w_k}
 \right],
 \\
 \boldsymbol z_{\rm orb}^{(n)}
 &=Q_n
 \left(\bigoplus_{\boldsymbol X\in\mathscr S_n}
 B_{\boldsymbol X}^{-1}\right)
 Q_n^{-1}\boldsymbol y_{\rm orb}^{(n)}.
 \end{aligned}
 \label{eq:reconstruction}
\end{equation}
Equation~\eqref{eq:reconstruction} is the central result.
As we announce below Eq.~\eqref{eq:one_inverse}, even if $M_{\rm orb}$ explodes as \(N_n\propto5^n/6\), we discover that inverting never takes more than $3\times3$.
A python notebook, where the reconstruction is implemented symbolically in SymPy
is available
in Ref.~\cite{mattiacci2026symb_calc}.

For measured data, Appendix~\ref{app:frequencies} explains how
to obtain \(\widehat{\boldsymbol y}_{\rm orb}^{(n)}\) from
event frequencies, including the outcome and orbit multiplicities.
Inserting this vector into Eq.~\eqref{eq:reconstruction} gives
estimates of the retained invariant combinations.
If a resource test uses only selected combinations,
its linear input map can be composed with
Eq.~\eqref{eq:reconstruction} in advance, so the remaining
invariant coordinates need not be estimated separately.
Finite-size entanglement certification from second- and third-order
invariants provides a concrete application~\cite{scala2026finite}.
This reduces the intermediate reconstruction steps; a comparison
of the measurement time needed for a prescribed precision requires
a separate statistical analysis.

\subsection{The bipartite case}
\label{sec:bipartite_case}
For two parties, the theorem gives a nine-dimensional
\(VV\) block \(H(d_A)\otimes H(d_B)\) and a one-dimensional
\(WW\) block \(\lambda(d_A)\lambda(d_B)\).
The transformation from the orbit basis to the sector basis goes as follows.

A vector $u \in \{e_{T_a}, e_{T_b}, e_{T_c}\}^{\otimes 2}$ is $S_3$ -invariant iff it is of the form 
\begin{align}
\alpha \sum_{i \in \{T_a, T_b, T_c\}} e_i \otimes e_i + \beta \sum_{i \ne j \in \{T_a, T_b, T_c\}} e_i \otimes e_j = 
\nonumber \\
\alpha (T=T) + \beta (T \ne T)
\end{align}
in the orbit basis. Hence the only $S_3$-invariant vector $u_2$ in $W \otimes W$ has $\alpha = -2\beta$ and the only $S_3$-invariant vector $u_1$ in $V \otimes V$ has $\alpha = \beta$. Summarising:
\begin{align}
    \left[ \begin{array}{l} u_1 \\ u_2 \end{array} \right]
    =
    \left[ \begin{array}{cc} 1 & 1 \\ 2 & -1 \end{array} \right]
    \left[ \begin{array}{c} T = T \\ T \ne T \end{array} \right]
\end{align}
and the coordinate transformation in this basis transition is described by the transposed matrix: 
\begin{equation}
 Q_{TT}=
 \begin{pmatrix}1&2\\1&-1\end{pmatrix},
 \qquad
 Q_{TT}^{-1}
 =\frac13
 \begin{pmatrix}1&2\\1&-1\end{pmatrix}.
 \label{eq:two_Q}
\end{equation}
All other orbit coordinates already match the \(VV\)
product coordinates, up to their ordering.
The two displayed directions have data coordinates
\((y_{T=T}+2y_{T\ne T})/3\) and
\((y_{T=T}-y_{T\ne T})/3\), respectively.
The same transformation applies to the merged input.

In particular, the scalar \(WW\) response gives
\begin{equation}
 y_{T=T}-y_{T\ne T}
 =
 \lambda(d_A)\lambda(d_B)(x_5-x_4).
 \label{eq:two_contrast_response}
\end{equation}
For a normalized state, this isolates the difference between
the global purity \(\operatorname{Tr}\rho^2\) and
\(\operatorname{Tr}[\rho(\rho_A\otimes\rho_B)]\).
The known bipartite nine-plus-one factorization~\cite{QST2025}
is thus the first instance of the general sector reconstruction.

\section{Partial transposition and information limits}
\label{sec:blindness}

Theorem~\ref{thm:main} reconstructs all combinations retained by the cycle-merging map
\(C^{\otimes n}\) (\ref{eq:cycle_sum}). To see why the discarded cycle orientations
cannot be separated, recall that the three probability factors
in Eq.~\eqref{eq:haar_data} share each local setting. After Haar
averaging, their local measurement coefficients obey
\begin{equation}
 [SW(d)]_{\pi,(123)}=[SW(d)]_{\pi,(132)},
 \qquad \pi\in\Pi.
 \label{eq:equal_cycle_coefficients}
\end{equation}
The two orientations therefore enter these equations only through
their sum. The map \(C\) records this information loss; inverting
the averaged equations cannot recover their difference.

Let \(\Gamma_j\) denote transposition on party \(j\).
A negative eigenvalue of \(\rho^{\Gamma_j}\) certifies
entanglement between that party and the rest~\cite{Peres1996};
its moments can also support entanglement tests~\cite{YuImaiGuhne2021}.
Partial transposition makes the lost cycle distinction explicit.
Moving it from the state to the permutation operator in
Eq.~\eqref{eq:contractions_general}, using
\(P_{\tau_j}^{T}=P_{\tau_j^{-1}}\), gives
\begin{equation}
 x_{\rho^{\Gamma_j}}(\tau_1,\ldots,\tau_n)
 =
 x_\rho(\tau_1,\ldots,\tau_j^{-1},\ldots,\tau_n).
 \label{eq:pt_contraction}
\end{equation}
Thus it exchanges the two three-cycles at party \(j\) alone;
the identity and transpositions are unchanged. This differs
from complex conjugation, which inverts every local label, and
common copy relabelling, which conjugates every label by the
same permutation.

Equation~\eqref{eq:pt_contraction} exchanges the two cycle
orientations at party \(j\), while \(C\) maps both to \(e_E\).
Consequently,
\begin{equation}
 \boldsymbol z^{(n)}(\rho^{\Gamma_j})
 =
 \boldsymbol z^{(n)}(\rho).
 \label{eq:pt_merged}
\end{equation}
The discarded local direction is
\(\ker C=\operatorname{span}\{e_{(123)}-e_{(132)}\}\);
\(\ker(C^{\otimes n})\) is spanned by tensors containing at
least one such factor. These directions are removed before
orbit reduction.

The same equality has a Born-rule explanation. Set
\(\boldsymbol U^{(j*)}
=(U_1,\ldots,\overline{U}_j,\ldots,U_n)\), where the bar
denotes entrywise conjugation. Equation~\eqref{eq:born} gives
\begin{equation}
 p_{\rho^{\Gamma_j}}(\boldsymbol i|\boldsymbol U)
 =
 p_\rho(\boldsymbol i|\boldsymbol U^{(j*)}).
 \label{eq:pt_probability}
\end{equation}
This relates generally different settings; it does not
equate the two expressions at the same \(\boldsymbol U\).
Applying Eq.~\eqref{eq:pt_probability} to each of the three
factors in Eq.~\eqref{eq:haar_data} gives, with
\(\mathrm d\boldsymbol U=\prod_k\mathrm dU_k\),
\begin{align}
 &\int\mathrm d\boldsymbol U\,
 \prod_{h=1}^3
 p_{\rho^{\Gamma_j}}(\boldsymbol i_h|\boldsymbol U)
 =
 \int\mathrm d\boldsymbol U\,
 \prod_{h=1}^3
 p_\rho(\boldsymbol i_h|\boldsymbol U^{(j*)})
 \nonumber\\
 &\quad=
 \int\mathrm d\boldsymbol U\,
 \prod_{h=1}^3
 p_\rho(\boldsymbol i_h|\boldsymbol U).
 \label{eq:pt_correlator}
\end{align}
The change of variables \(U_j\mapsto\overline{U}_j\) preserves
Haar measure and gives the last equality. Averaging therefore
removes the distinction between the two settings, in agreement
with Eq.~\eqref{eq:pt_merged}.
For two parties, partial transposition on \(B\) interchanges
\(x_9=\operatorname{Tr}\rho^3\) and
\(x_{10}=\operatorname{Tr}[(\rho^{\Gamma_B})^3]\).
Equation~\eqref{eq:cycle_pair_sum} retains only
\(2(x_9+x_{10})\), so these correlators cannot determine
\(x_9-x_{10}\) without additional information.

To access independently only $x_{10}$ one can define for all $R=(\boldsymbol i , \boldsymbol U)$ an unbiased estimator \(\widehat\rho(R)\), that does not need to be a density matrix\cite{Huang2020,Vermersch2024}, such that \(\mathbb E_{q_\rho}[\widehat\rho(R)]=\rho\) where
\(q_\rho(\mathrm d \boldsymbol U ,\boldsymbol{i})
=\mu(\mathrm d \boldsymbol U)p_\rho(\boldsymbol i \mid \boldsymbol U)\).
In this approach, each  shot $R$ retains both the random setting and the measured outcome~\cite{Elben2020}.
Therefore to access only $x_{10}$, the price to pay is an experiment where each applied setting $\boldsymbol U$ must be recorded.
Instead, in our protocol, we have access only to $x_9+x_{10}$, but we only require that the same \emph{unknown} $\boldsymbol U$ acting on each copy of $\rho$ is Haar-distributed.
Therefore the lack of information on $\boldsymbol{U}$ turns into an easier and more economical experimental procedure.  

In summary, the information limit
proved here is forced by the common-setting Haar averages and Theorem~\ref{thm:main} completes the
reconstruction of what survives those averages.

\section{Conclusion}\label{sec:conclusion}

We have determined which third-order local-unitary invariant
combinations can be reconstructed from products of three outcome
probabilities measured at a common local random setting and then
averaged. For any number of parties, the theorem recovers all
combinations retained by local cycle merging by accessing to the physical information with a symbolically invertion of $3\times 3$ matrices. This avoids to handle a numerical inversion of a matrix that explodes as the number of subsystems increases.
The protocol to access those physical quantities uses
independent local rotations, measurements on separately prepared
copies, and classical processing, making it compatible with
distributed experiments without the need of calibrating or storing each unitary.
The result distinguishes information present in these correlators
from information their equations cannot recover. In particular,
the local Haar averages force the bipartite third moments
\(\operatorname{Tr}\rho^3\) and
\(\operatorname{Tr}[(\rho^{\Gamma_B})^3]\) to appear only
through their sum.
The number of retained coordinates still grows exponentially, and
the theorem does not establish favorable statistical cost or
conditioning. Its practical appeal is that reconstruction does not
require recording the applied unitary matrices. Developing efficient
estimators for the retained combinations remains an open task, as
does extending the analysis to higher orders.

\section*{Acknowledgements}
We thank D. Chru\'{s}ci\'{n}ski, C. Lupo and M. Ziman for the fruitful scientific discussion. GSc is supported by grant NCN SONATA 21 UMO2025/59/D/ST2/02708. GSc thanks the Institute of Physics at the Slovak
Academy of Sciences for its hospitality through the SAIA
Support Fund.

\appendix

\section{Characters and dimensions of the invariant sectors}
\label{app:characters}

This appendix derives the dimensions of the invariant tensor sectors
and verifies that their sum equals the orbit count \(N_n\).
The required information is contained in the characters:
the traces of the representation matrices.

For a representation on \(\mathcal H\) with character \(\chi\),
averaging its matrices over the group projects onto
\(\mathcal H^{S_3}\). The average maps every vector to an invariant
vector and acts as the identity on vectors already invariant.
Its trace therefore equals the dimension of that subspace,
giving~\cite{serre1977linear}
\begin{equation}
 \dim\mathcal H^{S_3}
 =
 \frac16\sum_{\eta\in S_3}\chi(\eta)
 =
 \frac{\chi(1)+3\chi((12))+2\chi((123))}{6}.
 \label{app:eq:multiplicity}
\end{equation}
The weights count the group elements in each conjugacy class.
Thus invariant-subspace dimensions can be calculated without constructing
a basis of the subspace of invariant vectors.

The pattern action in Eq.~\eqref{eq:pattern_relabelling} gives
the local characters
\begin{equation}
 \begin{array}{c|ccc}
 &1&(12)&(123)\\
 \hline
 \text{class size}&1&3&2\\
 \chi_\gamma&5&3&2\\
 \chi_{W}&2&0&-1
 \end{array}
 \label{app:eq:character_table}
\end{equation}
The last row follows by subtracting the three invariant directions
in \(V\) from \(\chi_\gamma\).
Characters are constant on each
conjugacy class, so these values specify all six group elements.

Since \(\gamma\) is a permutation representation, its trace
counts local patterns invariant by the group element: the identity fixes all five,
a transposition fixes \(D,E\) and its matching \(T\) label,
and a three-cycle fixes only \(D,E\).
A tuple of local $\boldsymbol{\pi} \in \Pi^n$ is invariant under relabelling $\boldsymbol{\pi} \mapsto \eta \cdot \boldsymbol \pi, \forall \eta \in S_3$.
Equivalently, traces multiply under tensor products, so
\(\Gamma_n=\gamma^{\otimes n}\) has class character
\((5^n,3^n,2^n)\). Averaging with the class sizes \(1,3,2\)
in Eq.~\eqref{app:eq:multiplicity} gives
\begin{equation}
 N_n=|\Pi^n/S_3|
 =\frac{5^n+3\cdot3^n+2\cdot2^n}{6}
 =\dim(A^{\otimes n})^{S_3}.
 \label{eq:counts}
\end{equation}
This is also Burnside's orbit-counting formula.
It gives \(N_1=3\), \(N_2=10\), and \(N_3=37\), counting
orbit coordinates rather than $\left(\prod_{i=1}^n d_i\right)^2$ independent parameters of a
normalized physical state.

For \(k\ge1\), the character of \(W^{\otimes k}\) instead has
class values
\[
 \chi_{W^{\otimes k}}=(2^k,0,(-1)^k).
\]
We calculate the dimension of its invariant subspace using the formula (\ref{app:eq:multiplicity}). At \(k=0\), the empty tensor product is one-dimensional and
every group element acts as the identity, giving character
values \((1,1,1)\). Combining these cases yields
\begin{equation}
 w_k
 =
 \dim(W^{\otimes k})^{S_3}
 =
 \frac{2^k+3\delta_{k0}+2(-1)^k}{6}.
 \label{app:eq:wk}
\end{equation}

To verify completeness, sum the invariant-sector dimensions over
all choices of \(W\) positions. The invariant part of a sector
with \(k\) such positions has dimension \(3^{n-k}w_k\), and
there are \(\binom nk\) placements. The binomial theorem therefore gives
\begin{align}
 \sum_{k=0}^n\binom nk3^{n-k}w_k
 =
 \frac{(3+2)^n+3\cdot3^n+2(3-1)^n}{6}=N_n.
 \label{app:eq:dimension_sum}
\end{align}
The sector basis thus contains exactly the number of vectors
required by the orbit description.

Finally, Eq.~\eqref{app:eq:wk} vanishes only for \(k=1\).
Among the \(2^n\) choices of local \(V/W\) factors, precisely
\(n\) contain a single \(W\). Hence \(2^n-n\) sectors contribute
to the reconstruction.
\section{From measured frequencies to reconstruction coordinates}
\label{app:frequencies}

Equation~\eqref{eq:reconstruction} takes representative orbit
correlators \(\boldsymbol y_{\rm orb}^{(n)}\) as input.
Measured event frequencies instead count all outcome triples
belonging to a pattern orbit. This appendix converts those
frequencies into the coordinates required by the inverse map as in Eq.\eqref{eq:multiplicity}.

For an orbit \(o\subset\Pi^n\), the corresponding event contains every
outcome triple whose local equality-pattern tuple belongs to \(o\).
Choose a representative \(\boldsymbol\pi\in o\).
Using the local multiplicities \(m_\pi(d)\) introduced with
Eq.~\eqref{eq:normalization}, this representative has
\(\prod_{k=1}^n m_{\pi_k}(d_k)\) numerical outcome realizations.
Common copy relabelling preserves every local type, so the same
multiplicity applies to all \(|o|\) pattern tuples in the orbit.

All these realizations have the same averaged correlator
\(y_o=(\boldsymbol y_{\rm orb}^{(n)})_o\).
The probability mass of the complete orbit event is therefore
\begin{equation}
 P_o
 =
 |o|\left[\prod_{k=1}^n m_{\pi_k}(d_k)\right]y_o.
 \label{eq:orbit_probability}
\end{equation}
Here \(|o|\) counts pattern tuples, while the product counts
numerical outcome realizations of each tuple.

To estimate \(P_o\), form triples from three distinct measurement
shots at each product setting, as required by
Eq.~\eqref{eq:product}, and count the fraction belonging to
the event \(o\). 

For an orbit with nonzero outcome multiplicity, the
representative-coordinate estimate is
\begin{equation}
 \widehat y_o
 =
 \frac{P_o}
 {|o|\prod_{k=1}^n m_{\pi_k}(d_k)}.
 \label{eq:frequency_coordinates}
\end{equation}
Collecting these estimates in the prescribed orbit order gives
\(\widehat{\boldsymbol y}_{\rm orb}^{(n)}\), to which the
reconstruction formula~\eqref{eq:reconstruction} applies.

\end{document}